# 2E1 Ar$^{17+}$ decay and conventional radioactive sources to determine efficiency of semiconductor detectors

Emily Lamour[1], Christophe Prigent, Benjamin Eberhardt, Jean Pierre Rozet, and

Dominique Vernhet

Université Pierre et Marie Curie – Paris 6, UMR 7588, INSP, 140 rue de Lourmel, Paris,

F-75015 France, and

CNRS, UMR7588, INSP, Campus Boucicaut, Paris, F-75015 France

Although reliable models may predict the detection efficiency of semiconductor detectors, measurements are needed to check the parameters supplied by the manufacturers namely the thicknesses of dead layer, beryllium window and crystal active area. The efficiency of three silicon detectors has been precisely investigated in their entire photon energy range of detection. In the 0 to a few keV range, we have developed a new method based on the detection of the 2$E$1 decay of the metastable Ar$^{17+}$ 2s→1s transition. Very good theoretical knowledge of the energetic distribution of the 2$E$1 decay mode enables precise characterization of the absorbing layers in front of the detectors. In the high-energy range (> 10 keV), the detector crystal thickness plays a major role in the detection efficiency and has been determined using a $^{241}$Am source.

---

[1] Corresponding author. <emily.lamour@insp.jussieu.fr>

# I. INTRODUCTION

Due to their excellent compromise between resolution and detection efficiency, semiconductor detectors (Ge, Ge(Li), Si(Li), Si, HPGe and so on) are commonly used for quantitative analysis of X-ray emission. Model of their response function and detection efficiency is entirely based on the knowledge of photon-matter processes, *i.e.*, mainly the photoelectric effect and the Compton scattering, and allows to reach a precision of a few % in absolute X-ray yield measurements. Being easy to handle, these solid-state detector have brought, for several decades, meaningful results in various fields of interest from particle to solid-state physics. In our group, we routinely employed detectors with Si(Li) and Si crystals, either to determine absolute populations of excited states of highly charged ions produced during ion-solid interaction[1,2], or to quantify the production rates of X-ray emission when rare gas clusters are submitted to strong optical fields[3].

The overall accuracy of these measurements strongly relies on both the response function and detection efficiency of those detectors. Nowadays, the response functions are well understood and simulations, frequently based on Monte Carlo methods, reproduce well all the spectral features whatever the incident energy[4,5,6,7,8]. Below a few keV, the low energy tail accompanying the full absorption energy peak is known to be mainly due to incomplete charge collection effects[5,6,7] that occur at low incident photon energies. For much higher incident photon energies (above a few tens of keV), a small fraction of photons interacts with the quasi-free electrons of the detector material and is scattered, losing part of their energy in the detector or escaping from the detector. These Compton processes induce a continuum background[4, 8] under the main peaks, which result from the photoelectric effect. The energy-efficiency relationships have been largely studied both experimentally and theoretically[9, 10, 11, 12, 13]. The most common experimental techniques exploit either various calibrated radioactive samples providing X-ray and γ lines from 3 to 60 keV[4, 9, 13], or monochromatic X-ray synchrotron radiation from 0.1 to a few keV[6, 7] provided by storage rings.

In this paper, we present a new approach to determine the detection efficiency in the low energy range. It is based on the accurate theoretical knowledge of the line shape of the single-photon energy distribution of the two-photon decay mode (2$E$1) of the Ar$^{17+}$ 2$s$ metastable state, which exhibits a broad energetic distribution from 0 to 3 keV[14]. This measurement enables the determination of absorbing layers in front of the active area of detectors such as beryllium or organic windows, metallic contacts and dead layers when present. Moreover, the thickness of the active crystal area of detectors has been deduced from measured detection efficiencies above 10 keV, using a $^{241}$Am source. Measurements for three different solid-state detectors used have been performed: a Si(Li) detector from ORTEC Inc. (USA) and two silicon drift detectors, a XFlash model from RONTEC GmbH (Germany) and the other supplied by KETEK GmbH (Germany).

The paper is organized as follows. In section II, specifications of each detector are presented and basic points to calculate their total transmission are reviewed. In section III, we present the experimental methods to determine the thickness of different components which play a significant role in the detection efficiency. Below 10 keV, we make use of the deexitation of highly charged ions interacting with gaseous and solid targets as well as fluorescence emission induced by electron impact onto various solid foils (section II.A.). Above 10 keV, we report on results obtained with a $^{241}$Am radioactive source (section II.B). The last section summarizes the comparison of detector characteristics extracted from experimental data with those supplied by the manufacturers. We discuss the entire set of results presenting the full efficiency curve over the whole energy range of detection. In conclusion, the sensitivity of the two methods used is emphasized.

## II. DETECTOR CHARACTERISTICS AND CALCULATION OF DETECTION EFFICIENCIES

### A. Detector characteristics

The specifications given by each manufacturer are summarized in table I. All the three semiconductor detectors can be used under good vacuum conditions (~ $10^{-7}$ mbar) and are sealed by a beryllium window. The Si(Li) detector from ORTEC Inc. (USA) is equipped with the thickest crystal giving access to efficiencies larger than 10% up to 70 keV while the 25 μm Be window thickness limits the detection efficiency below 1 keV. A crystal diameter of 10 mm insures a large solid angle. Other specific features are a gold layer acting as a front electrode and a thin dead layer of 0.1 μm. Finally, the ensemble is mounted on a liquid nitrogen cryostat for cooling purpose. With this first generation technique, a 180 eV FWHM resolution at 5.9 keV is obtained. The two other detectors are based on SDD (silicon drift detector) technique[15], which combines a large sensitive area (the entire Si wafer is sensitive to radiations) with a small output capacitance due to a sub-millimetre crystal thickness. They both can afford high counting rates and reach a FWHM resolution of 140 eV at 5.9 keV, which is significantly better than for the ORTEC detector. A Peltier element cools them down to a working temperature of around − 10 °C. The RONTEC XFlash detector has a thinner crystal (300 µm) combined to the thinnest Be window (8 µm). This leads to a detection range (efficiency better than 10%) extending from 850 eV to 25 keV. The KETEK detector (most recent generation of SDD detectors), equipped with a slightly thicker crystal (450 µm) and a beryllium window of 25 µm, has a detection range of 1 to 30 keV. In addition, its much larger diameter, *i.e.*, 10 mm instead of 2 mm for the RONTEC detector, insures a much better solid angle. In the following, the different set of specifications provided by the manufacturers for each detector will be introduced as parameters in the calculation and compared to our measurements.

**B. Calculation of the total transmission of a semiconductor detector**

The absolute detection transmission T is simply given by the product of the solid angle ($\Omega/4\pi$) with the quantum efficiency $\varepsilon(E)$:

$$T = \varepsilon(E) \times \frac{\Omega}{4\pi}. \tag{1}$$

The quantum efficiency corresponds to the ratio of detected photons over the number of incident photons and depends drastically upon the incident photon energy (E): when raising the X-ray energy, the efficiency first increases to reach a maximum, and then decreases reaching "zero" when the X-ray energy is so high that radiation crosses the active volume without interaction. The following simple formula reproduces this behavior:

$$\varepsilon(E) = (1 - \exp[-\mu_{sz}(E) \times x_{sz}]) \times \exp\left[-\sum_i (\mu_i \times x_i)\right], \quad (2)$$

The first term corresponds to the intrinsic efficiency of the detector, where $\mu_{sz}$ stands for the photoelectric absorption coefficient in the sensitive zone of crystal thickness $x_{sz}$ (generally $x_{sz}$ is not well known *a priori*). $\mu_i$ and $x_i$ denote respectively the absorption coefficients and thicknesses of the different absorbers lying in front of the active region of the crystal such as the Be window ($\mu_{Be}$, $x_{Be}$), the Si dead layer ($\mu_{dl}$, $x_{dl}$) and/or the gold layer ($\mu_{Au}$, $x_{Au}$). Usually, an additional correction factor has to be considered $F_{escape} = (1 - P_{escape})$ that corresponds to the escape peak correction where $P_{escape}$ stands for the probability to get such event. This peak appears when a $K_\alpha$ fluorescence photon, first created by photoelectric effect, escapes from the active area. The incident photon with an initial energy E will be then detected at the energy $E-E_K$ where $E_K$ is the K-shell binding energy of the crystal constituent[9]. $P_{escape}$ is given by:

$$P_{escape} = \frac{1}{2} \times \frac{\mu_E^K}{\mu_E} \times \omega_K \times \left[1 - \frac{\mu_S}{\mu_E} \ln\left(1 + \frac{\mu_E}{\mu_S}\right)\right]. \tag{3}$$

$\mu_E^K$ is the K-shell photoelectric absorption coefficient. $\mu_E$ and $\mu_S$ stand for the total absorption coefficients of the incident photons and the $K_\alpha$ X-rays emitted by crystal atoms respectively; $\omega_K$ is the K-shell fluorescence yield. For Si or Si(Li) detectors, $F_{escape}$ gives rise to a correction of the detection efficiency smaller than 2% while it can reach more than 10% for Ge crystals.

It is worth to recall that detection efficiency is very sensitive to absorber thicknesses for incident photon energies less than 10 keV. Especially, the beryllium thickness affects strongly the efficiency in the 0 to 3 keV energy range while the dead layer and gold contact thicknesses play a major role between 2 and 6 keV. For photon energies greater than 10 keV, within the Be window thickness range considered here, only the thickness of the sensitive zone may dramatically change the efficiency. Consequently, we have developed two methods to experimentally test the efficiency of our three silicon detectors: one sensitive to the characteristics of absorbers lying in front of the active area and the other allowing us to determine the crystal thickness.

## III. METHODS OF MEASUREMENT AND RESULTS BELOW AND ABOVE 10 keV

**A. Low energy region below 10 keV**

Precise measurements of the detection efficiency in the low energy range (< 3 keV) have been obtained by recording photons coming from the deexcitation of the $Ar^{17+}$ 2s metastable state. Briefly summarized, a beam-foil spectroscopy experiment has been performed on the LISE (Ligne d'Ions Super Epluchés) facility at GANIL (Grand Accélérateur National d'Ions Lourds - Caen, France)[1, 2]. Fully stripped 13.6 MeV/amu $Ar^{18+}$ ions, directed onto thin solid carbon targets of a few µg/cm² and gaseous targets ($N_2$ and $CH_4$) lead to the production of excited $Ar^{17+}$ ions by the single capture process. Most of the populated exited states decay very fast via single photon modes towards the ground state, *i.e.*, within a few $10^{-14}$ s for the n*p* states (the cross section of the capture process being maximum in *n* = 2). On the contrary the 2*s* state has a long lifetime (3.5 $\times 10^{-9}$ s for $Ar^{17+}$). At a projectile velocity of 13.6 MeV/amu, this lifetime corresponds to a propagation distance of 173.6 mm. Consequently, the deexcitation of such a long lifetime state may be easily recorded by a solid-state detector placed at 90° with respect to the beam axis and at a distance of a few centimetres behind the target. A specific collimation system, mounted in front of the detector, ensures good spatial resolution and allows us to know precisely the detection solid angle.

The 2*s* state has two decay modes: a two-photon mode (2*E*1) and a single-photon magnetic mode (*M*1) with branching ratios respectively of 97% and 3%[14] for $Ar^{17+}$. In the 2*E*1 decay mode, two photons are simultaneously emitted sharing the 2*s*→1*s* transition energy (E(2*s*→1*s*) = 3.321 keV in the projectile frame, *i.e.*, 3.273 keV in the laboratory frame). For such medium Z ion, the well known energy distribution[14] is then found to be a broad continuum from 0 to E(2s→1s) and symmetric with respect to half of the total energy E(2s→1s). Consequently, this transition provides a way to test the detection efficiency continuously at very low energy by recording a single spectrum. More specifically, it affords to achieve precise measurements of the Be entrance window of the detector. Figure 1 shows that the 2E1 energy distribution is dramatically changed when varying the Be thickness while the sensitivity to the dead layer appears to play a role above 1.839 keV, *i.e.*, the Si K-shell threshold energy.

We have analyzed 35 spectra recorded by the RONTEC detector placed at a 50 mm distance behind gaseous targets and solid foils of different thicknesses[2]. Precise determination of the experimental 2E1 line shape has been achieved since only total counts change from one spectrum to the other. Figure 2 presents a typical spectrum obtained for a $N_2$ gaseous target. We can easily distinguish the 2E1 photon decay emission from the M1 line. Additionally, a small peak at 1.74 keV is visible, coming from the $K_\alpha$ Si fluorescence. Together with the experimental spectrum, we present plots taking into account the calculated efficiency using equation 2 and the convolution of the theoretical energy distribution with detector resolution (see Fig. 2). Those fits, which have been calculated for a given dead layer and different thicknesses of the entrance Be window, show clearly that a 8 μm Be thickness, as specified by the company, is too thin to reproduce the recorded emission. The best fit of the 2E1 line below 1.7 keV is obtained with a Be thickness of 15 ± 1 μm and an unexpected dead layer of 0.07 μm to account for the observed Si fluorescence peak (no dead layer has been specified by the manufacturer see table I). To illustrate this effect, we present in Fig. 3 a comparison with and without dead layer for the same Be window thickness (namely 15 µm). The introduction of a Si dead layer thickness, even very small, visibly improves the agreement with photon emission spectrum recorded by the RONTEC detector. We stress the fact that, the whole range of energies between 0 and 3 keV being covered at once with the 2$E$1 spectral distribution, unprecedented accuracies can be obtained compared to methods using a set of monochromatic lines.

These results have been confirmed by measurements of fluorescence yield induced by electron impact. A 10 keV electron beam is directed at 90° onto targets of different elements; *i.e.*, foils or powders of $NaNO_3$, $MgF_2$, KClAl, Si, $CaF_2$, Sc and stainless steel (CrFeNi). The production of vacancies in atomic inner-shells of these various components gives rise to the generation of characteristic X-rays. The well known full efficiency calibration of one detector allows then to calibrate others quite rapidly. Two X-ray detectors, among which one is calibrated, are symmetrically placed at 30° from the electron beam direction to record the emitted X-rays. Circular diaphragms of well-defined diameter[16] are positioned in front of the detectors to monitor the solid angle $\Omega$. The whole experimental set-up is under good vacuum ($< 10^{-6}$ mbar). Since one of the detectors (det0) is fully calibrated in efficiency, the detection efficiency of the other detector (det1) is simply given by:

$$\varepsilon_{det1} = \frac{N_{det1}^{K\alpha}}{N_{det0}^{K\alpha}} \times \frac{\Omega_{det0}}{\Omega_{det1}} \times \varepsilon_{det0} \quad (4)$$

where $N_{det1}^{K\alpha}$ and $N_{det0}^{K\alpha}$ stand for the number of counts (corrected by the acquisition dead time of the corresponding channel) of a given $K_\alpha$ line recorded by each detector. Beside crossed check measurements of the efficiency of the RONTEC detector, a full calibration of the KETEK detector has then been achieved. Moreover, we have performed measurements of the Mn $K_\alpha$ line emitted by a $^{55}$Fe source to enlarge the number of experimental data. As shown in Fig. 4, the experimental efficiencies measured by this method in the 0-6 keV range are fully consistent with those obtained *via* the 2*E*1 energy distribution; although they are not so accurate (the error bars for X-ray energy above 2 keV are much larger). Nevertheless, the fluorescence measurements give access to the thickness of the Be window of the KETEK detector, which is found larger than expected (34 µm) with an uncertainty less than 10%.

Same complete studies had been performed previously with the ORTEC detector (see spectra in Ref 1) and the specifications provided by the manufacturer were found to be much more accurate as summarized at the end of the paper (see table III).

### B. High energy region above 10 keV

The detection efficiency in the photon energy regime above 10 keV is sensitive only to the thickness of the crystal active area ($x_{sz}$) as discussed in section II. In this region, the equation (2) simplifies to $\varepsilon(E) = 1 - \exp[-\mu_{sz}(E) \times x_{sz}]$. Evaluation of crystal thickness can be merely obtained by comparing, on the same spectrum, the number of counts of two peaks at different energies, provided relative emission probabilities of the corresponding transitions are well established. We have set a radioactive $^{241}$Am source in front of each detector and recorded the $^{237}$Np L X lines and γ-rays whose energies range from 10 to 60 keV, coming from alpha decay of $^{241}$Am atoms. The interest of using such a source lies in the precise knowledge of emission probabilities of almost 20 transitions[17, 18, 19]. In practice, the recorded spectrum is characterized by four groups of Np X$_L$ emission lines, *i.e.*, L$_l$, L$_α$, L$_β$ and L$_γ$ and four main gamma rays as shown in Fig 5 where a spectrum recorded by the ORTEC detector is displayed. Since the crystal active area is much thinner for the RONTEC and KETEK detectors, only the first γ-ray at 26.35 keV is observed with enough statistics and the contribution of Compton scattering background is also much lower. In Fig. 6, this background has been subtracted under each group of lines using a linear fit so as to extract the photoelectric efficiency from the intensity.

In table II, transition energies and corresponding branching ratios (I$_{Line}$) normalized to the L$_α$ line are reported. In order to reduce the statistical error bars, the L$_α$ line has been chosen as reference since it is the most intense peak recorded by our three detectors. The relative detection efficiency ($\varepsilon_{Line}/\varepsilon_{L\alpha}$) can be determined through the following simple equation:

$$\frac{\varepsilon_{Line}}{\varepsilon_{L_\alpha}} = \frac{N_{Line}}{N_{L_\alpha}} \times \frac{I_{L_\alpha}}{I_{Line}}. \tag{5}$$

where N$_{Line}$ and N$_{L\alpha}$ are the number of counts of each line.

An enlargement of the 10 to 25 keV energy region is presented in Figure 6 for the ORTEC and the RONTEC detectors allowing us to illustrate the strong dependence of the ratio N$_{line}$/N$_{L\alpha}$ with the type of detector due to the difference of crystal thickness. We can clearly see for instance that the detection efficiency of the $L_\beta$ line is much higher for the ORTEC than for the RONTEC detector.

The experimental ($\varepsilon_{Line}/\varepsilon_{L\alpha}$) values obtained by this method and the corresponding fitted crystal thickness ($x_{sz}$) are presented in table II. It is worth mentioning that gamma rays with energies higher than 30 keV are observable only on the spectrum recorded with the ORTEC detector (see Fig 5) since for the others, the detection efficiency is around 4% at 30 keV reaching almost "zero" above 60 keV.

Outcomes of this method are illustrated in figure 7 where experimental results are compared to fits obtained either with a crystal thickness that matches experimental data, or the one given by the company in the case of the RONTEC and ORTEC detectors. For the ORTEC detector no difference is found within the error bars confirming the characteristics provided by the manufacturer. For the RONTEC detector, the best fit is obtained for a crystal thickness of 240 μm, which is significantly lower than the value given by the manufacturer. Finally, the same methods have been applied to determine the efficiency of the KETEK model reported as well in table III and good agreement is found in this case with the crystal thickness specified by the manufacturer.

## IV. FULL EFFICIENCY DETERMINATION AND CONCLUSION

With the entire set of data presented above, we can summarize the few discrepancies found with the parameters provided by the manufacturers by comparing the thicknesses of each element playing a role in the detection efficiency of the three solid-state detectors investigated here (see Table III). For the RONTEC and KETEK detectors, the beryllium window thickness is found to be systematically larger than specified; a disagreement of a factor of two is almost reached in the case of the RONTEC detector. Moreover, it seems that the introduction of a very thin dead layer ($70 \pm 20$ nm) reproduces better the experimental $2E1$ spectrum in the case of the RONTEC, for which no dead layer was expected. Although special emphasis has been put on the minimization of the effective silicon dead layer by the manufacturer, the slight reduction of the efficiency might also be due to a non-structured $p^+$ junction cover of the entrance window (see Ref 15 for the detector description). For the KETEK detector, we can only give an upper limit value for the dead layer thickness since only efficiency measurements deduced from fluorescence x-ray yields have been performed. Regarding the crystal thickness, the experimental value found for the RONTEC detector is 20% lower than the manufacturer's data while good agreement within the error bars is obtained for the others. Figure 8 summarizes the complete detection efficiency for the three solid-state detectors.

In conclusion, we clearly demonstrate that such complete experimental studies are needed to check parameters provided by the manufacturers. Indeed, reliable values of the detection efficiency within a few % of uncertainty are mandatory to extract *absolute* cross sections of numerous processes giving rise to soft or hard X-ray emission such as those occurring during short laser pulses interacting with rare gas clusters, highly charged ions colliding with atoms, clusters, surfaces or solids or even to characterize plasma density and temperature in an ECR ion source. We have measured the efficiency of three silicon detectors among which two are based on the silicon drift detector technique. We showed that the detection of the Ar$^{17+}$ 2s→1s transition and, especially, of the 2$E$1 decay mode emission is a successful method to determine the detection efficiency at low photon energy. In fact, the precise theoretical knowledge of the 2$E$1 energetic distribution offers the opportunity to accurately investigate the efficiency from 0 to around 3 keV and provides a stringent test of both, the beryllium window thickness and possible very thin absorber layers lying in front of the sensitive crystal area. Finally, the use of a $^{241}$Am source, which gives rise to photons in the 10-60 keV range, enables a precise evaluation of the effective thickness of the detection crystal.

**Acknowledgment**

This work has been supported by the French national agency of research (ANR: Agence Nationale de Recherche) contract ANR 06 BLAN 0233. We thank the CIRIL platform of the CIMAP laboratory and the staff at GANIL for providing us with outstanding technical assistance and high-quality of highly-charged ion beams.

**TABLE I. Specifications of each detector provided by the manufacturers.**

| Detector | crystal | | dead layer | contact layer | Be thickness | Energy resolution |
|---|---|---|---|---|---|---|
| | Thickness | Diameter | | | | |
| RONTEC | 300 μm | 2 mm | 0 | -- | 8 μm | 140 eV |
| KETEK | 450 μm | 10 mm | 0 | -- | 25 μm | 140 eV |
| ORTEC | 5.53 mm | 10 mm | 0.1 μm | 0.02 μm | 25 μm | 180 eV |

**TABLE II.** Energies, branching ratios for the Np $X_L$ lines and $\gamma$-rays from the alpha decay of a $^{241}$Am source and the relative detection efficiency deduced from eq(5) for the three detectors. The branching ratios are normalized to the $L_\alpha$ line (having an emission probability of 0.130 per disintegration); the reported values are consistent with those published in refs 17, 18, 19 with an accuracy of ±5%. $L_\alpha = L_{\alpha 1} + L_{\alpha 2}$, $L_\beta(a) = L_{\beta 15} + L_{\beta 2} + L_{\beta 7}$ and $L_\beta(b) = L_{\beta 5} + L_{\beta 1} + L_{\beta 3}$. The fitted value of the sensitive zone of the crystal detector is reported on the last line.

| Line | Energy (keV) | Normalized branching ratio ($I_{Line}$) | RONTEC model | KETEK model | ORTEC model |
|---|---|---|---|---|---|
| | | | $\left(\varepsilon_{Line}/\varepsilon_{L_\alpha}\right)_{exp.}$ | | |
| $L_l$ | 11.89 | $6.0\times10^{-2}$ | 1.42±0.08 | 1.28±0.07 | 1.02±0.05 |
| $L_\alpha$ | 13.90 | 1 | 1±0.05 | 1±0.05 | 1±0.05 |
| $L_\beta(a)$ | 16.91 | $3.53\times10^{-1}$ | 0.67±0.04 | 0.74±0.04 | 1.02±0.05 |
| $L_\beta(b)$ | 17.76 | 1.02 | 0.56±0.04 | 0.67±0.03 | 1.02±0.05 |
| $L_\gamma$ | 20.79 | $3.75\times10^{-1}$ | 0.32±0.02 | 0.45±0.02 | -- |
| $\gamma_1$ | 26.35 | $1.72\times10^{-1}$ | 0.20±0.01 | 0.23±0.01 | 0.89±0.05 |
| $\gamma_2$ | 33.20 | $9.04\times10^{-3}$ | -- | | 0.63±0.05 |
| $\gamma_3$ | 43.42 | $4.80\times10^{-3}$ | -- | | 0.37±0.02 |
| $\gamma_5$ | 59.54 | 2.66 | -- | | 0.16±0.01 |
| Fitted crystal thickness | | | $x_{sz} = 240\pm1$ µm | $x_{sz} = 450\pm2$ µm | $x_{sz} = 5755\pm350$ µm |

**TABLE III. Thickness of each element playing a role in the calculation of the detection efficiency for the three detectors studied: exp.value for the experimental value extracted from the measurements; man.spec. for the specifications provided by the manufacturer. For the KETEK detector, just an upper value of the dead layer (*) is reported since it has been deduced only from fluorescence measurements (see text). It is worth mentioning that a gold layer has to be taken into account for the ORTEC Si(Li) detector and is found to be equal to 0.020 ± 0.005 μm.**

| Detector | Be thickness $x_{Be}$ (μm) | | dead layer $x_{dl}$ (μm) | | sensitive zone $x_{sz}$ (μm) | |
|---|---|---|---|---|---|---|
| | exp.value | man.spec. | exp.value | man.spec. | exp.value | man.spec. |
| RONTEC | 15±1 | 8 | 0.07±0.02 | 0 | 240±1 | 300 |
| KETEK | 34±3 | 25 | < 0.05$^{(*)}$ | 0 | 450±2 | 450 |
| ORTEC | 26.5±0.5 | 25 | 0.10±0.03 | 0.1 | 5755±350 | 5530 |

FIGURE CAPTIONS

FIG 1. (Color online) Energy distribution of the 2$E$1 decay mode for Ar$^{17+}$. The theoretical distribution: dotted line; the expected experimental distributions for three cases (two different Be windows and dead layer, respectively $x_{Be}$ et $x_{dl}$): dash-dotted line with $x_{Be}$ = 8 µm and $x_{dl}$ = 0, dashed line with $x_{Be}$ = 15 µm and $x_{dl}$ = 0 and red solid line $x_{Be}$ = 15 µm and $x_{dl}$ = 0.07 µm. Note that the experimental distributions take into account the convolution with the detector resolution.

FIG 2. (Color online ) Spectra of the 2$s$→1$s$ transition recorded by the RONTEC detector placed at 50 mm behind the target (green thin solid line). Fits have been obtained using a dead layer ($x_{dl}$) of 0.07 µm and three different Be window thicknesses ($x_{Be}$) to calculate the detection efficiency with formula (2).

FIG 3. (Color online) Spectra of the 2$s$→1$s$ transition recorded by the RONTEC detector placed at 50 mm behind the target (green thin solid line). Fits have been obtained using a unique Be window thickness of 15 µm and two different dead layer thicknesses ($x_{dl}$) to calculate the detection efficiency with formula (2).

FIG 4. (Color online) Efficiency of the RONTEC detector in the 0-6 keV photon energy range: dashed line corresponds to the expected efficiency according to the manufacturer's specifications; green solid line with error bars is the fitted curve

obtained with eq(2) for $x_{Be} = 15\pm1$ μm and $x_{dl} = 0.07\pm0.03$ μm that have been determined *via* the 2*E*1 energy distribution (see text and Fig. 2 and 3); circles symbolize the measured efficiencies using fluorescence emission.

FIG 5. (Color online) Typical spectrum recorded by the ORTEC detector showing the Np $X_L$ lines and γ -rays coming from alpha decay of a [241]Am source. Also clearly visible is the "shelf" due to Compton scattering.

FIG 6. (Color online) Spectrum of the 10 to 25 keV energy region recorded with a [241]Am placed in front of the ORTEC (blue) and RONTEC (green) detectors. Note that the Compton scattering contribution has been subtracted in both cases.

FIG 7. (Color online) Efficiency of two different detectors (RONTEC and ORTEC models) for photon energy > 10 keV. For the RONTEC detector (in green), the circles symbolize the experimental efficiency values, the solid line corresponds to the best fit obtained with $x_{sz} = 240$ μm; the black dashed line is the efficiency calculated with the manufacturer's crystal thickness. For the ORTEC detector (in blue), the square symbols are the experimental data and the dash-dotted line the best fit, which is in agreement with the manufacturer specification.

FIG 8. (Color online) Efficiency over the entire energy range of detection based on the different experimental methods for the three detectors under investigation.

**Symbols correspond to the experimental data: circles for the RONTEC detector (in green); triangles for the KETEK detector (in red) and squares for the ORTEC one (in blue). Lines display the best fit calculated with the parameters extracted from experiments and given in Table III.**